# Reservoir Computing using High Order Synchronization of Coupled Oscillators

**A A Velichko[1,3], D V Ryabokon[2], S D Khanin[2], A V Sidorenko[1], A G Rikkiev[1]**

[1]Institute of Physics and Technology, Petrozavodsk State University, 33 Lenin str., 185910, Petrozavodsk, Russia
[2]Russia Marshal of the Soviet Union Budennyi Military Academy of Communications, 3 Tikhoretsky pr., 194064 St. Petersburg, Russia

[3] E-mail: velichko@petrsu.ru

**Abstract**. We propose a concept for reservoir computing on oscillators using the high-order synchronization effect. The reservoir output is presented in the form of oscillator synchronization metrics: fractional high-order synchronization value and synchronization efficiency, expressed as a percentage. Using two coupled relaxation oscillators built on $VO_2$ switches, we created an oscillator reservoir that allows simulating the XOR operation. The reservoir can operate as with static input data (power currents, coupling forces), as with dynamic data in the form of spike sequences. Having a small number of oscillators and significant non-linearity, the reservoir expresses a wide range of dynamic states. The proposed computing concept can be implemented on oscillators of diverse nature.

## 1. Introduction

Neural networks are effective tools for solving problems of pattern recognition and artificial intelligence [1]. One of the promising areas in the development of neural networks is reservoir computing systems (RC) ) [2–5].

A typical architecture for reservoir calculations consists of the following parts (Figure 1):
1. An input single-layer neural network has a matrix of weights $W_I$ and reflects the input data to the internal nodes of the reservoir.
2. A recurrent layer (reservoir) performs non-linear transformation of the input data.
3. The readout layer is a trained neural network with a matrix of weights $W_R$, and it converts the output of the reservoir into the resulting response of the entire system.

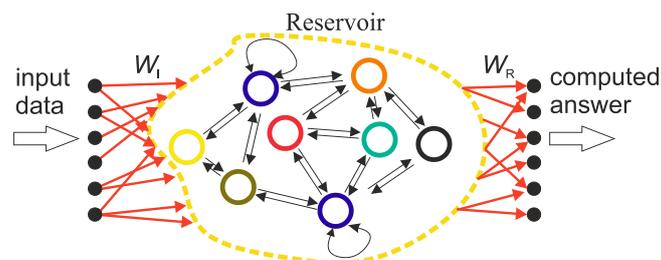

**Figure 1.** Typical architecture of reservoir computer.

The main task of the reservoir is to transfer the input data into a higher dimensional space that the output neural network can classify the result more accurately. A similar technique in the machine learning is called a kernel trick [6,7]. The main difference between a reservoir computer (RC) and recurrent neural networks is the learning process. In RC, the input connections $W_I$ and the connections in the reservoir are set randomly and do not change. Only the coefficients of the readout layer $W_R$ are trained. It creates advantage of significant reduction of RC training [2].

Widely used reservoir systems include reservoirs based on nonlinear dynamic systems (oscillators [3,8,9], memory elements [4]) and reservoirs with signal delay effects [10]. The coupled oscillators exhibit rich dynamic behavior [11], including synchronization, clustering, and the formation of chimeric states that can be used in RC. Endeavours to develop reservoir calculations have been attempted on oscillators of various physical nature: mechanical oscillators [9], chemical [8], spin-torque nano-oscillators [12], phase oscillators [13]. Variety of existing oscillator systems determines the need of developing a new concept of reservoir computing, proposed in this study, which can be applied to oscillators of any nature.

To demonstrate the possibilities of neural network to solve complex problems, a nonlinear XOR operation can be used as a test example. The applications to solve it include the use of the dynamic properties of two coupled oscillators supplemented by an adder and a peak detector [14], the concept of oscillatory threshold logic [15], spike neural network [16] and memristor networks [17].

In this study, we propose a new concept of the oscillator reservoir, which operation is based on the high-order synchronization effect and special synchronization metrics. To illustrate the operation of a reservoir, we implement the XOR operation.

## 2. High order synchronization of coupled oscillators reservoir

Oscillators of any type (phase, relaxation, etc.) and any nature (electrical, mechanical, magnetic, etc.) possess a synchronization effect [11]. From a practical perspective, an interesting and important phenomenon of high-order synchronization can be explained by the following example. Two coupled relaxation oscillators generate spike pulses, for example, current or voltage, with own frequencies $F^0_1$ and $F^0_2$ (see Figure 2a). The coupling coefficients of the oscillators are $\Delta_{1,2}$ and $\Delta_{2,1}$. These parameters determine the intensity how the pulse of one oscillator tries to initiate or suppress the pulse of another oscillator. A synchronization effect may occur, when oscillator pulses appear synchronously with a certain synchronization pattern. On a Figure 2b, the pulses of two oscillators at certain points in time are in phase, and the peaks synchronize through the number of periods $M_1$ and $M_2$. As a result, the steady frequencies differ from the own frequencies $F_1 \neq F^0_1$ and $F_2 \neq F^0$, and the steady frequencies are related to each other as integers $F_1 : F_2 = M_2 : M_1$. This illustrates the high-order synchronization effect, when the oscillators are synchronized at high subharmonics in their spectra.

Earlier [18], we proposed a family of metrics consisting of two parameters — the high-order synchronization value SHR and the synchronization efficiency value $\mu$.

If the system contains N oscillators, then the synchronization value can be determined between any two oscillators with numbers $i, j$:

$$\text{SHR}_{i,j} = M_j : M_i, \tag{1}$$

where $M_i$ and $M_j$ are the number of periods between synchronous spikes. Figure 2b demonstrates examples of synchronization with $\text{SHR}_{1,2}=3:2$, $\text{SHR}_{1,2}=1:1$ and $\text{SHR}_{1,2}=1:2$.

The value of the synchronization efficiency $\mu_{i,j}$ determines what percentage of the signal has a given $\text{SHR}_{i,j}$. A signal is considered synchronized if $\mu_{i,j}$ exceeds a predetermined threshold value $\mu_{i,j} \geq \mu_{th}$ (for example, $\mu_t = 90\%$). SHR and $\mu$ are time-integrated parameters that are calculated on a temporal waveform containing a significant (50-10000) number of oscillations. A more detailed calculation procedure can be found in [18], and a calculation accounting for chimeric synchronization is elaborated in [19].

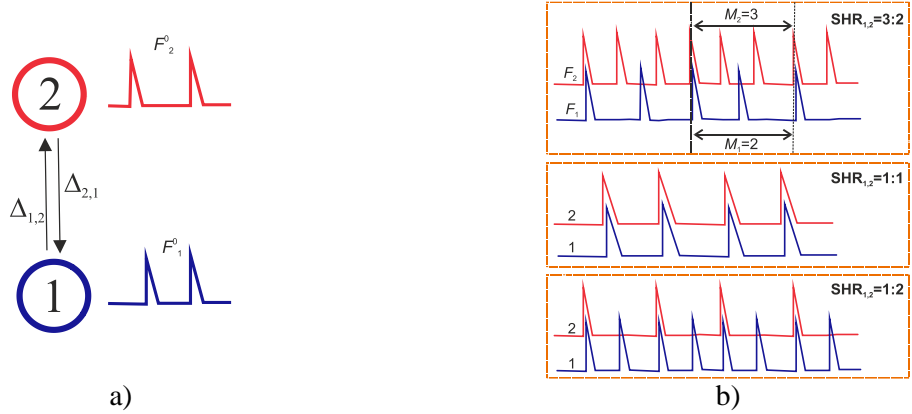

**Figure 2.** Designation of two coupled oscillators (a). Examples of high-order synchronization of two oscillators with $SHR_{1,2}=3:2$, $SHR_{1,2}=1:1$ and $SHR_{1,2}=1:2$.

As an oscillator model, we take a relaxation oscillator circuit based on a switch with an S-shaped I–V characteristic (Figure 3a). Oscillations occur when the capacitor $C$ is periodically discharged through a switch in the presence of a noise source $U_{in}$. The capacitor is charged by the current source $I_p$. The value of the oscillation frequency $F^0$ is a complex function of the parameters of the circuit and the switch. We used switching elements based on vanadium dioxide, which can be obtained in laboratory conditions [20] and have high operation speed due to the metal–semiconductor phase transition [21]. The physics of the switching effect allows the implementation of thermal coupling of oscillators on $VO_2$ structures. We described the thermal coupling model in details in [20,22]. The model is based on the effect, when one of the switches goes into the on state, it emits temperature waves, which reduce the threshold switching voltage $U_{th}$ of the neighboring switch by $\Delta$.

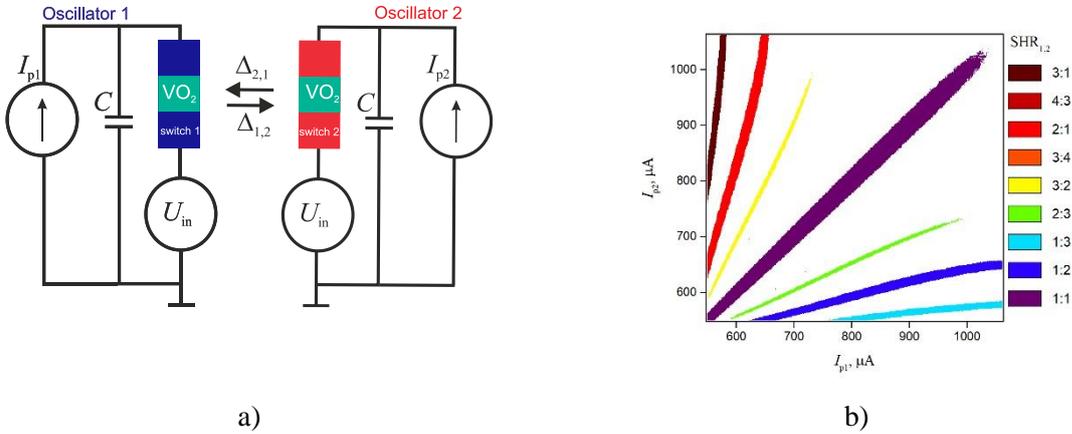

**Figure 3.** Circuit of relaxation oscillators based on a $VO_2$ switching element with thermal coupling ($I_p$ is the current source, $C$ is the capacitance, $U_{in}$ is the noise source, $\Delta_{1,2}$ and $\Delta_{2,1}$ are the coupling coefficients) (a). The distribution of the high-order synchronization value $SHR_{1,2}$ of two oscillators in the space of the supply currents $I_p$ (Arnold's tongue) (b).

The oscillation frequency can be varied by the supply currents of the circuit. Figure 2b presents an example of the $SHR_{1,2}$ synchronization distribution for a circuit of two oscillators, when the supply currents $I_{p1}$ and $I_{p2}$ are varied (the parameters of the circuit and calculations are described in [23]). The synchronization distribution has the shape of Arnold's tongue. On the diagonal, where $I_{p1} \approx I_{p2}$, we observe synchronization at the main harmonics $SHR_{1,2}=1:1$. Above the diagonal, synchronization has the value $SHR_{1,2}>1$, and, below the diagonal, we observe $SHR_{1,2}<1$.

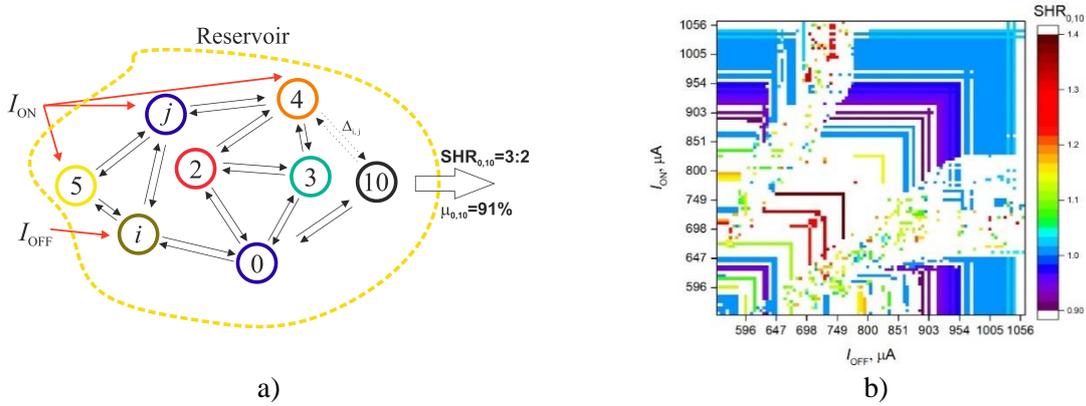

a)                              b)

**Figure 4.** A reservoir of coupled oscillators with input and output parameters (a). The distribution of $SHR_{1,10}$ for a system of 10 coupled oscillators taken from [18] (b).

A more complex picture of $SHR_{i,j}$ distribution is observed in the case of a large network of coupled oscillators (Figure 4a). Figure 4b illustrates $SHR_{1,10}$ distribution for a system of 10 coupled oscillators taken from [18]. Two levels of supply currents $I_p = I_{ON}$ and $I_p = I_{OFF}$, distributed between oscillators, are fed to the network input. The distribution pattern of $SHR_{1,10}$ has a non-linear shape, with synchronization regions alternating in a complex manner. The type of distribution depends on the pattern of supply currents and connection levels.

In this way, the system of oscillators serves as a reservoir where the input signal is transformed nonlinearly into synchronization metrics $SHR_{i,j}$ and $\mu_{i,j}$. A similar system can be used for reservoir computing.

### 3. Implementation of the XOR operation

In linear classification, all input samples can be divided into corresponding classes by a straight line - for the samples of dimension two, by a plane - for dimension three, or by a hyperplane - for $n$-dimensional samples. For classification, one neuron with $n$ inputs and a certain weight at each input is sufficient. If, for division into classes, several straight lines or hyperplanes are needed, then the problem is nonlinear. The XOR operation is nonlinear classification problem that converts a pair of binary input values to 0 or 1 according to the rule presented in Table 1.

**Table 1.** XOR definition.

| Input | | Output |
|---|---|---|
| *X* | *Y* | *Q* |
| 1 | 1 | 0 |
| 1 | 0 | 1 |
| 0 | 1 | 1 |
| 0 | 0 | 0 |

The XOR operation has two input parameters, which can be displayed on the plane, and the output values can be represented as colored figures (see Figure 5a). The output values belong to two classes (1 or 0), and, to separate them on the plane, at least two straight lines (line 1, line 2) are needed. The implementation of the XOR operation requires at least three neurons, one neuron for each line, and a third neuron combines the result. The Figure 5b illustrates displacement neurons, as well as arrays of weights $W_I$ and $W_R$ between three layers (input, hidden and output).

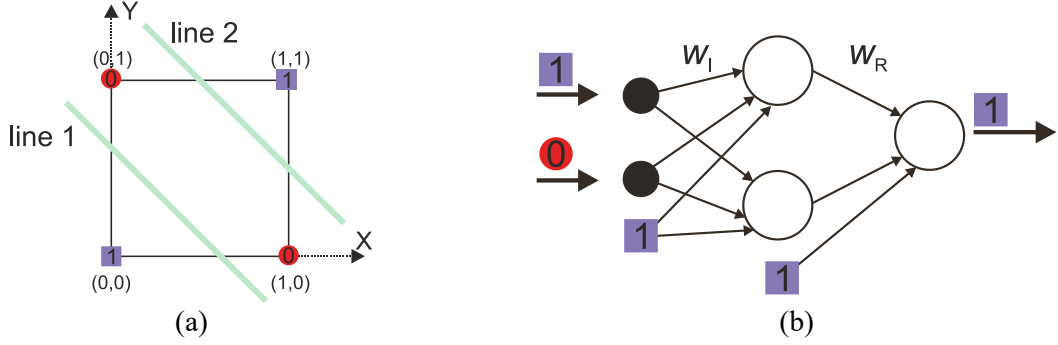

**Figure 5.** XOR operation on a plane, in the space of X and Y parameters, with dividing lines (a). Neural network for XOR operation (b).

An alternative way to divide input samples into classes is to introduce the third parameter Z, which transfers a two-dimensional problem into a three-dimensional one. Figure 6a demonstrates a conversion of the XOR task into a three-dimensional view, where a plane divides the output values into two classes. In this way, a nonlinear problem is transformed into a linear one, which requires only one neuron. To increase the dimension of the problem, and to generate an additional parameter Z, reservoir networks are used [2], which non-linearly transform the input parameters. As a result, the number of neurons and connections decreases and a reservoir appears in the network (Figure 6b).

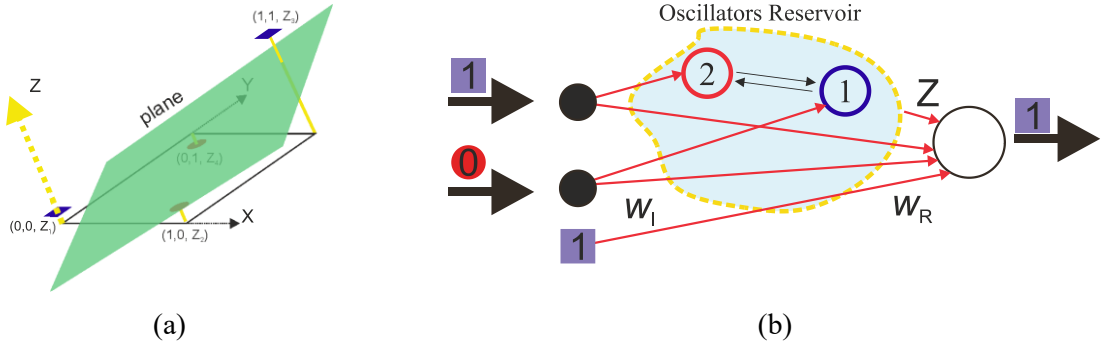

**Figure 6.** XOR operation in 3-dimensional space of parameters X, Y, Z, with the dividing plane (a). Neural network for XOR operation with one neuron and oscillatory reservoir (b).

For solving the XOR problem, the two coupled oscillators considered above can be used as a reservoir, and the high-order synchronization value $SHR_{1,2}$ serves as parameter Z. To approximate the XOR function using two oscillators, we select the following dependences and parameters of the neuron. The values of the inputs are linearly converted into the values of the oscillator currents according to the equations:

$$I_{p1} = 638\mu A + 343\mu A \cdot X$$
$$I_{p2} = 574\mu A + 416\mu A \cdot Y$$
(2)

$SHR_{1,2}$ values are determined by the Figure 3b distribution depending on current values.
The activation function of the output neuron is a threshold function:

$$Q = f(\Sigma) = \begin{cases} 1, \text{if } \Sigma < 0 \\ 0, \text{if } \Sigma \geq 0 \end{cases},$$
(3)

where $\Sigma$ is the sum of the combined input of the output neuron, determined by the equation:

$$\Sigma = 1 \cdot (1.12) + X \cdot (-0.8) + Y \cdot (0.78) + SHR_{1,2} \cdot (-1).$$
(4)

Equation (4) already has the weights of neuron connections $W_R$. The current values $I_{p1}$ and $I_{p2}$, $SHR_{1,2}$, $\Sigma$ and $f(\Sigma)$ are listed in Table 2. The shape of the parameter space and the separating plane are presented in Figure 7.

**Table 2.** Input and output parameters of the reservoir and neural network when simulating XOR operation.

| X | Y | $I_{p1}$, µA | $I_{p2}$, µA | $SHR_{1,2}$ | $\Sigma$ | $Q=f(\Sigma)$ |
|---|---|---|---|---|---|---|
| 1 | 1 | 981 | 990 | 1:1=1 | 0,10 | 0 |
| 1 | 0 | 981 | 574 | 1:3=0.33(3) | -0,01 | 1 |
| 0 | 1 | 638 | 990 | 2:1=2 | -0,10 | 1 |
| 0 | 0 | 638 | 574 | 2:3=0.66(6) | 0,45 | 0 |

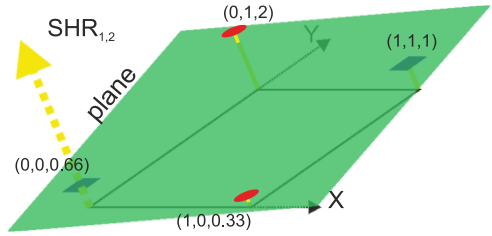

**Figure 7.** XOR operation in 3-dimensional space.

The $SHR_{1,2}$ values in Table 2 vary widely and nonlinearly. It makes possible to separate the two classes with a plane and get the correct value for the XOR operation.

## 4. Discussion and conclusion

As SHR takes various discrete values for certain parameters of the oscillator circuit, and non-synchronized and non-synchronized regions alternate nonlinearly on the parameter plane, the same circuit can recognize multilevel input data and be a basis for the implementation of complex logic circuits. The absence of synchronization state ($\mu_{i,j} < \mu_{th}$) can also be used by setting $SHR_{1,2}=0$.

The reservoir on the oscillators can have two types of output parameters: the synchronization values $SHR_{i,j}$ and the synchronization efficiency $\mu_{i,j}$. For the implementation of XOR operation, we used only $SHR_{i,j}$.

In addition to the oscillator currents, the coupling values can also serve as input parameters in the reservoir. For example, currents ($I_1$ and $I_2$) and the connections between the oscillators ($\Delta_{1\rightarrow 2}$ and $\Delta_{2\rightarrow 1}$) can be used as four input parameters for two oscillators. For a network of $N$ fully connected oscillators, the number of couplings is $N \cdot (N-1)$, and the total number of possible input parameters is the square of the number of oscillators $N^2$. Therefore, 28 oscillators are needed to operate with input data in the form of images, which have dimension of 28 by 28 pixels (MNIST database) and 728 pixels.

The number of synchronous states $N_s$ of the reservoir depends on the magnitude of noise, coupling strengths, and the number of oscillators [19,23]. The more $N_s$ are available, the more options we have for separating hyperplanes in the parameter space and the probability of finding a solution to the classification problem. Therefore, the reservoir capacity $N_s$ directly affects the network's ability to classify objects. The relative location of the synchronization areas and the order of their alternation with areas with no synchronization also influence the result. In a multi-oscillator system, the influencing factors include the choice of the pair of oscillators to take the output signal from, and the choice of oscillators and coupling forces to which the input signal is supplied.

Input data can enter the reservoir not only in the form of levels of static values (supply currents $I_p$, couping strengths $\Delta_{i,j}$), but also in the form of dynamic signals (spike sequences and harmonic signals). When an external periodic or quasiperiodic signal affect individual oscillators, the output values of $SHR_{i,j}$ and $\mu_{i,j}$ will change. Therefore, the system allows the operation with a wide range of data types.

The nonlinearity of the output data is caused by the complex nonlinear dynamics of the oscillators interaction, and by the way the value $SHR_{i,j}$ is set in the form of a fractional ratio of integers.

The reservoir computing based on oscillators and a high-order synchronization effect can be implemented on oscillators of diverse nature (electrical, magnetic, optical, biological, etc.). Therefore, the presented concept can find wide practical application in the field of creating artificial intelligence systems and requires further academic input.

**Acknowledgment**
This research was supported by the Russian Science Foundation (grant no. 16-19-00135).